\documentclass[final,5p,times]{elsarticle}

\usepackage{graphicx}

\usepackage{amssymb}

\usepackage{xspace}
\usepackage{textcomp}

\usepackage{booktabs}
\usepackage{graphicx}
\usepackage{caption}
\usepackage{subcaption}
\usepackage{hyperref}
\usepackage{siunitx}

\newcommand{\tlu}{TLU\xspace}

\newcommand{\tp}{\textsc{TelePix}\xspace}
\newcommand{\pmos}{{p-mos}\xspace}
\newcommand{\nmos}{{n-mos}\xspace}
\newcommand{\hb}{\textsc{HitBus}\xspace}
\newcommand{\corry}{\textsc{Corryvreckan}\xspace}
\makeatletter
\def\ps@pprintTitle{%
	\let\@oddhead\@empty
	\let\@evenhead\@empty
	\def\@oddfoot{}%
	\let\@evenfoot\@oddfoot}

\makeatother
\journal{Nuclear Instruments and Methods A}

\begin{document}
	\begin{frontmatter}
		
		\title{{Upgrading the Beam Telescopes at the DESY II Test Beam Facility}}
		
		\author[a]{H.~Augustin}		
		\author[d]{R.~Diener}
		\author[a]{S.~Dittmeier}
		\author[b]{P.~M.~Freeman}
		\author[c]{J.~Hammerich}
		\author[d]{A.~Herkert}
		\author[d]{L.~Huth\corref{cor1}}
		\ead{lennart.huth@desy.de}
		\author[a]{D.~Immig}
		\author[d]{U.~Kr\"amer}
		\author[d]{N.~Meyners}
		\author[e]{I. Peri\'{c}}
		\author[d]{O.~Sch\"afer}
		\author[a]{A.~Sch\"oning}
		\author[d]{A.~Simancas}
		\author[d]{M.~Stanitzki}
		\author[b]{D. Stuart}
		\author[a]{B.~Weinl\"ader}
		\cortext[cor1]{corresponding author}

		
		\address[a]{Physikalisches Institut der Universit\"at Heidelberg, INF 226, 69120 Heidelberg, Germany}
		\address[b]{Department of Physics, University of California, Santa Barbara, CA 93106, USA}
		\address[c]{University of Liverpool, Liverpool L69 3BX, United Kingdom}
		\address[d]{Deutsches Elektronen-Synchrotron DESY, Notkestr. 85, 22607 Hamburg, Germany}
	    \address[e]{Institut f\"ur Prozessdatenverarbeitung und Elektronik, KIT, Hermann-von-Helmholtz-Platz 1, 76344 Eggenstein-Leopoldshafen, Germany}
	

		\begin{abstract}
						The DESY II Test Beam Facility is a key infrastructure for modern high energy physics detector development, providing particles with a small momentum spread in a range from \SIrange[range-units=single]{1}{6}{\GeV} to user groups e.g. from the LHC 
						experiments and Belle II as well as generic detector R\&D. Beam telescopes are provided in all three test beam areas as precise tracking reference without time stamping, with triggered readout and a readout time of $>$~\SI{115}{\micro\second}. If the highest available rates are used, multiple particles are traversing the telescopes within one readout frame, thus creating ambiguities that cannot be resolved without additional timing layers.  Several upgrades are currently investigated and tested: 
						Firstly, a fast monolithic pixel sensor, the \tp, to provide precise track timing and triggering on a region of interest is proposed to overcome this limitation. The \tp is a \SI{180}{\nano\meter} HV-CMOS sensor that has been developed jointly by DESY, KIT and the University of Heidelberg and designed at KIT. In this publication, the performance evaluation is presented: The difference between two amplifier designs is evaluated. A high hit detection efficiency of above \SI{99.9}{\percent} combined with a time resolution of below \SI{4}{ns} at negligible pixel noise rates is determined. Finally, the digital hit output to provide region of interest triggering is evaluated and shows a short absolute delay with respect to a traditional trigger scintillator as well as an excellent time resolution.
						Secondly, a fast LGAD plane has been proposed to provide a time resolution of a few 10 ps, which is foreseen to drastically improve the timing performance of the telescope. Time resolutions of below 70~ps have been determined in collaboration with the University of California, Santa Barbara.
			
		\end{abstract}
	\end{frontmatter}

	\tableofcontents
	\section{Introduction}
	The development of novel particle detectors, be it for example upgrades of the LHC experiments, Belle II or more generic research in the context of  next generation high energy physics experiments, relies crucially on tests under realistic conditions. Test beam facilities like the one at DESY~II~\cite{DIENER2019265} are optimally suited for this purpose.
	The characterization of detector prototypes in a test beam requires precise knowledge of 4D track information, which is typically provided by beam telescopes.
	At the DESY~II Test Beam Facility, three EUDET-type pixel beam telescopes~\cite{jansen2016} provide an unprecedented pointing resolution of down to \SI{2}{\micro\meter}.
	This is achieved by the small pitch (ca. \SI{18}{\micro\meter}) of the MIMOSA-26~\cite{baudot2009,huguo2010} pixel sensors and an ultra-low material budget of  \SI{0.07}{\percent}~of a radiation length per tracking layer.
	Each of the six sensors is read out in a rolling shutter mode with the duration of one readout cycle of \SI{115}{\micro\second}.
	A \tlu~\cite{Baesso_2019} generates triggers based on a scintillator coincidence and exchanges trigger numbers with connected devices, allowing for data synchronization on hardware level.
	The maximum trigger rate that can be processed by the system is \SI{3}{\kilo\hertz} while the available particle rates can go up to several \SI{10}{\kilo\hertz}. Therefore, data frames with multiple hits per sensor creating ambiguities will occur.\\
	The main drawback of the MIMOSA-26 sensors is the lack of additional hit time information. 
	Extending the EUDET-type telescopes by an additional pixelated sensor that adds a precise time stamp to the spatial information allows for timing studies of devices under test (DUTs). This is crucial, e.g.\ to study if hits can be reliably assigned to the correct LHC bunch crossing. In early prototyping phases, the active area of a DUT is often small compared to the size of the MIMOSA-26 sensors (\SI[parse-numbers=false]{2x1}{\centi\meter\squared}). 
	In this case, an additional pixelated layer can also resolve the above-mentioned ambiguities by implementing a configurable region-of-interest (ROI) trigger output, that rejects particles outside the DUTs acceptance. This allows for efficient data taking with low rate DAQ systems.\\
	None of the already available sensors at the test beam combine a fine enough time stamp with trigger capabilities. In addition, they represent a significant amount of material when put in the beam, due to their hybrid design.\\   
	A test chip, the \tp, in a \SI{180}{nm} HV-CMOS technology is developed to overcome these limitations, by providing a time stamp with a precision below \SI{5}{ns}, combined with a fast digital hit output signal allowing for a configurable region of interest (ROI) trigger.\\
	
	Another approach that is currently under investigation is the use of low gain avalanche diodes (LGADs) to provide timing with a precision of a few \SI{10}{ps}, where single channels can also be used as a coarse ROI trigger for the telescope readout.
	
	\section{\tp as a Timing and Triggering layer}
	\subsection{Sensor Design}
	The \tp chip layout is based on previous submissions, which are described elsewhere \cite{SCHIMASSEK2021164812,PERIC2007876}.\\ 
	It compromises \SI[parse-numbers=false]{29x124}{} pixels with a pitch of \SI[parse-numbers=false]{165x25}{\micro\meter\squared}. 
	\tp is part of a full reticle submission with several test chips that has been received in late autumn 2020. 
	Signals are amplified inside the active pixel and digitized in the periphery at one edge of the sensor. 
	Both the time-of-arrival (ToA, \SI{10}{bit}) and time-over-threshold,(ToT, \SI{10}{bit}) are registered in the periphery. 
	The ToA is sampled on both clock edges, which doubles the timestamp precision without increasing the required clock frequencies.
	\tp implements two different amplifiers with a \pmos ( \nmos ) input transistor for columns 0-14 (15-29). 
	Each part has its own global externally applied threshold.  The sensor can be biased to up to \SI{-80}{V} to create a thick active depletion layer.
	Each pixel can be switched off individually, while the top half of the sensor features a 3 bit threshold trimming. 
	The chip operates trigger-less and continuously streams out data at 1.25~Gbit/s. 
	Finally, \tp features a fast digital trigger output option with a configurable pixel column range called \hb.\\
	
	\subsection{Test Beam characterization}

\begin{figure}[htbp]
		\centering
		\includegraphics[width=.8\columnwidth]{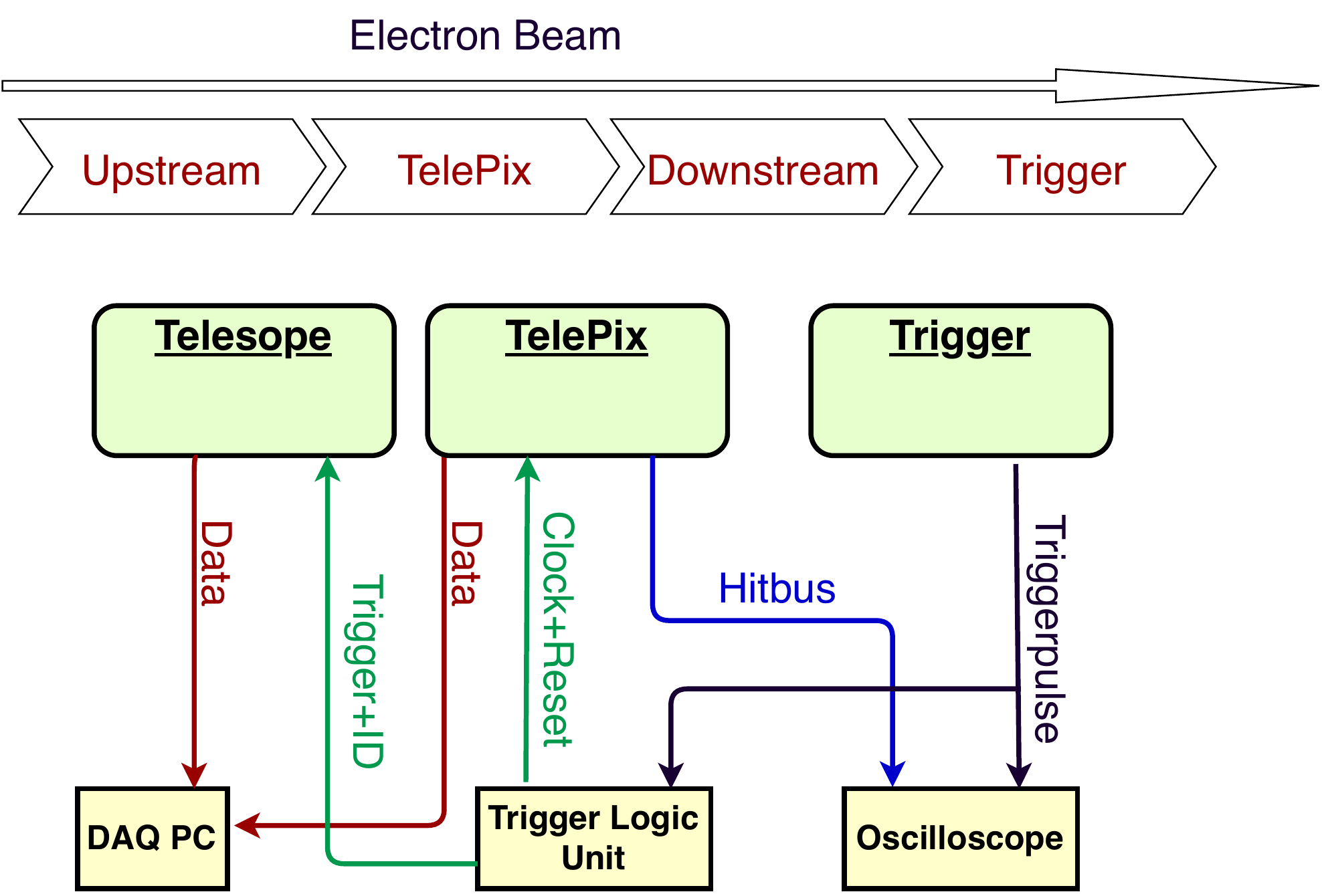}
		\caption{\label{fig:setupTB} Test setup during the test beam campaign. For triggering tests, the scintillator and digital hit out are connected to a 4 GHz oscilloscope. }
	\end{figure}

	The existing DAQ framework \cite{Dittmeier2018,TWEPP2017} of the Mu3e collaboration is used to configure and read out \tp. 
	All investigation results shown in the following are obtained at the DESY~II test beam facility. The sensor bias is set to \SI{-70}{V} if not stated otherwise.
	Figure~\ref{fig:setupTB} depicts the setup used to characterize the sensor in beam. The \tp is placed between two triplets of a provided reference telescope, which is framed by two scintillators generating triggers to activate the telescopes' readout. Telescope and \tp are synchronized with an AIDA-TLU~\cite{Baesso_2019}: The telescope exchanges trigger IDs with the TLU, while the \tp DAQ receives a phase stable \SI{125}{MHz} clock from the TLU and a reset signal at run start. The reset signal sets all counters in \tp to zero. Automated data taking is realized via a full integration into EUDAQ2\cite{Liu:2019wim}.\\
	The recorded data is analyzed with \corry~ \cite{Dannheim_2021}: Around each \tlu trigger, a time window with $\pm$\SI{10}{\micro\second} is created and all \tp hits with matching timestamps are added to the frame.
	Subsequently, clustering in space (and time for the \tp) is performed. 
	The particle trajectories are reconstructed based on the General-Broken-Lines formalism~\cite{gbl}, with the timestamp of the track being defined by the trigger time stamp from the \tlu.
	Only events with a single reference trajectory are kept for analysis, reducing the available data by a few percent, while providing a very clean sample. 
	Clusters on the \tp, that are within \SI{200}{\micro\meter} around the particle intersection point and $\pm$ \SI{500}{ns} around the track timestamp are assigned to the tracks and kept for further analysis. A constant time offset of the \tp is subtracted in the search window.\\
	
			\begin{figure}[htbp]
		\centering
		\includegraphics[width=.8\columnwidth]{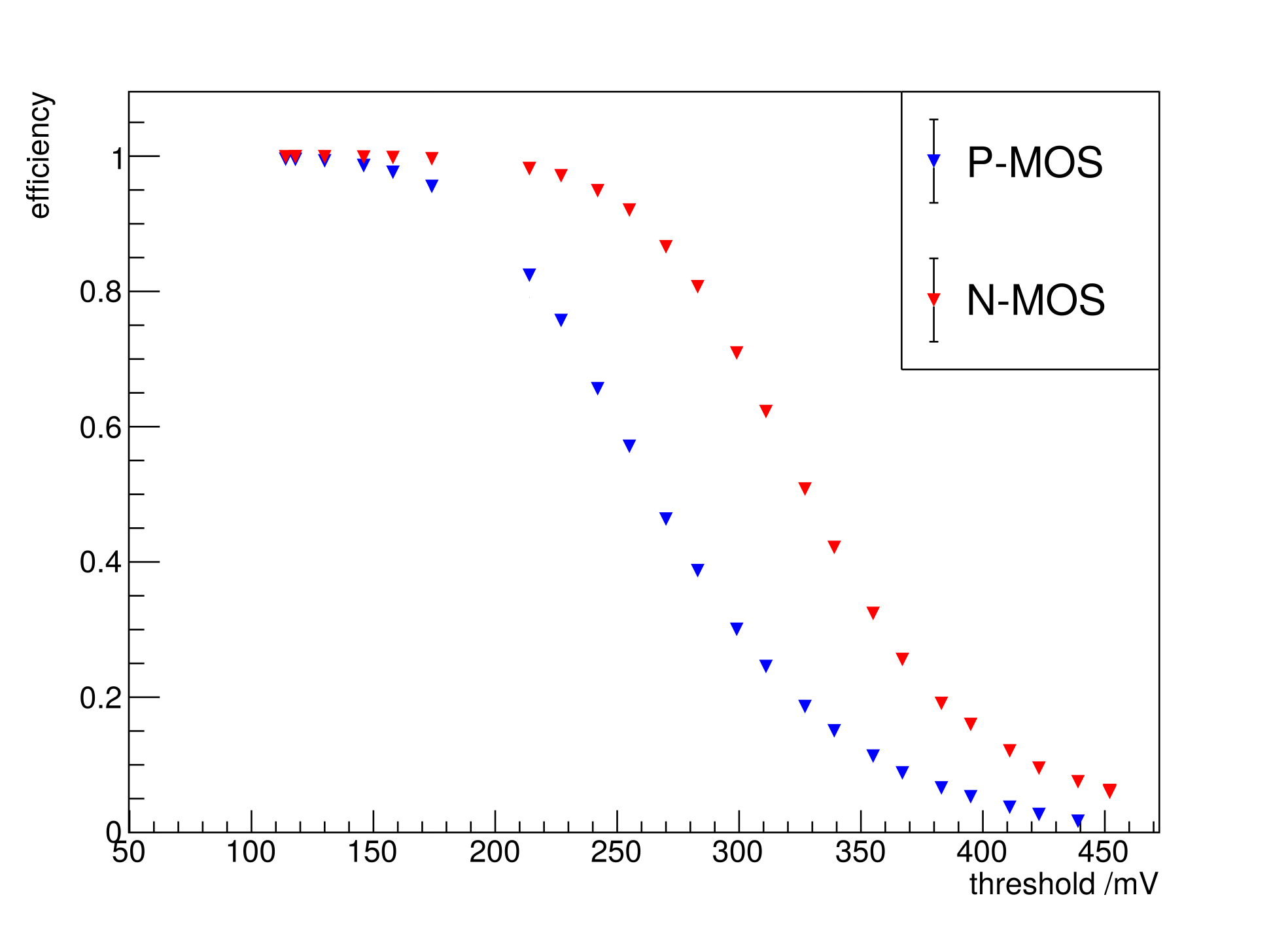}
		\caption{\label{fig:effScan} Hit detection efficiency as function of the detection threshold for the \pmos (blue) and \nmos (red) amplifiers. Note that statistical errors are included, but too small to be visible.A bias of \SI{-70}{V} is applied}
	\end{figure}

	\subsubsection*{Hit Detection Efficiency}
	The hit detection efficiency is defined as the number of tracks, that have an assigned cluster over the total number of reconstructed tracks, that intersect the DUT.  Figure~\ref{fig:effScan} shows the efficiency as a function of the externally applied detection threshold. Below \SI{100}{mV} thresholds, the readout saturates due to high rates of noise induced pixel hits. Detection efficiencies of above \SI{99.9}{\percent} are determined for the lowest thresholds reached in this study. After a highly efficient region, that is significantly larger for the \nmos type, efficiency decreases firstly at the corners and edges due to charge sharing, that causes less available charge per individual pixel. For even higher thresholds, the efficiency also decreases in the center of the pixels.

	\subsection*{Time Resolution}
\begin{figure}[htbp]
\centering
\includegraphics[width=.8\columnwidth]{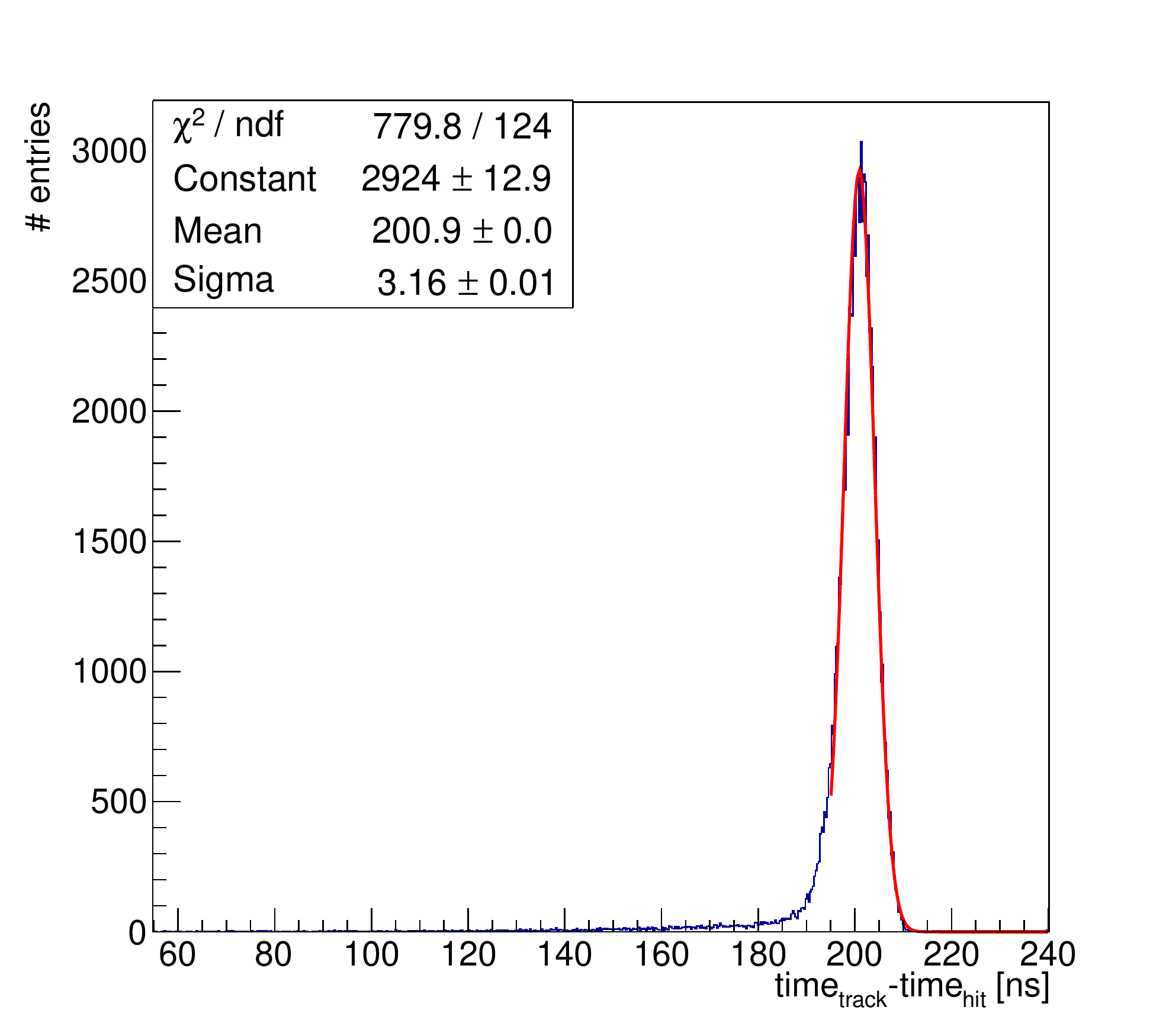}
\caption{\label{fig:timeRes} Exemplary time resolution of the \nmos amplifier section at a detection threshold of  \SI{108}{mV}. The tail towards the left is introduced by charge sharing and can be assigned to  clusters with a width of two.}
\end{figure}

	The time difference to the trigger scintillators is shown in figure~\ref{fig:timeRes} for a low threshold of \SI{108}{mV}. A fit to the distribution's core results in a time resolution of \SI{3.16\pm0.01}{ns}, without correcting for the time resolution of the scintillator signal. The entries in the tail of the distribution are explained by events in pixel corners with cluster size $>$ 1, with significant charge sharing and therefore less charge per pixel causing a longer delay for crossing the detection threshold, so called time walk.
	
	\subsection{Region of Interest Triggering Capabilities}
\begin{figure}[htbp]
		\centering
		\includegraphics[width=.8\columnwidth]{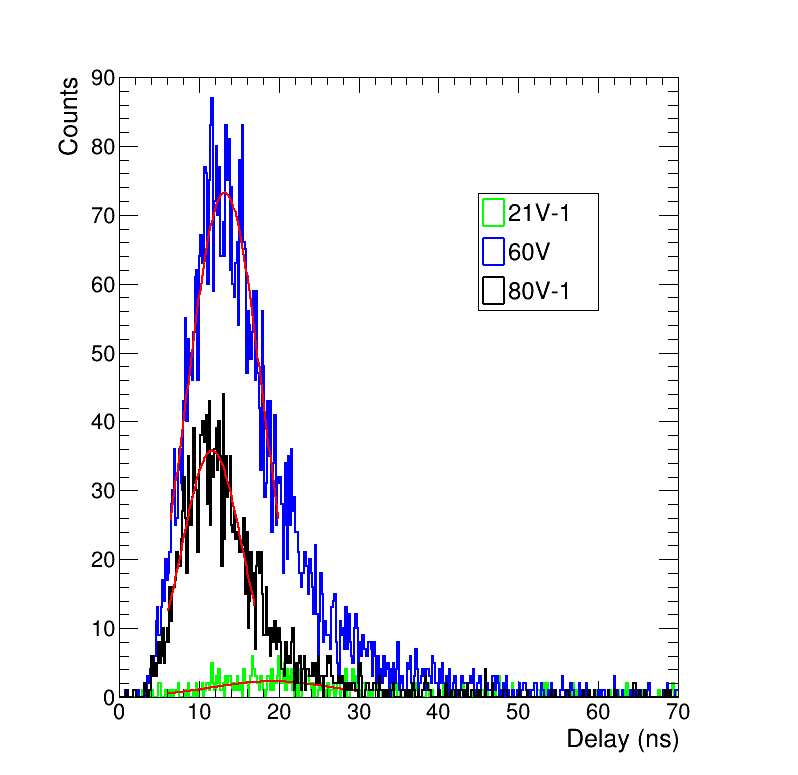}
		\caption{\label{fig:hb_delay}The measured time resolution and delay of the fast hit-or signal relative to a trigger scintillator coincidence with \SI{4}{\GeV} electrons.}
	\end{figure}

	The trigger capabilities have been studied with a similar setup as depicted in figure~\ref{fig:setupTB}. A \SI[parse-numbers=false]{2x1}{cm\squared} plastic scintillator, read out via a PMT is placed closely behind the \tp. Both, the \hb and the PMT, are connected to an oscilloscope, which  is triggered on the \hb and the absolute latency between the leading edges of the \hb and scintillator pulse is histogrammed directly with  an oscilloscope. Due to a design flaw, the fast logic OR can only be used in columns without any pixels being masked. Hence, a high threshold of \SI{151}{mV} is used to suppress any effect from noisy pixels.
	Figure \ref{fig:hb_delay} shows the results for three bias voltages, with a Gaussian fitted to the core of the distributions while table~\ref{tab:delays} summarizes the results.
	Delays (Resolutions) from \SI[parse-numbers=false]{22.55 (3.81)}{ns} to \SI[parse-numbers=false]{28.04 (6.56)}{ns} are measured and do agree with the expected behaviour for increasing bias voltage: the average signal size increases as the depletion region grows and less fluctuations on the threshold crossing time and a shorter signal delay are the consequence. In addition, the detector capacity is reduced for higher bias creating less noise.	
	All measured delays are acceptably short to be feasible as a trigger input for the \tlu, which takes approximately \SI{180}{ns} to process a trigger. The time resolution is consistent with the measurements presented above, if the higher threshold is taken into account.
	
	\begin{table}
		\centering
		\begin{tabular} {r c l}
			\toprule
			Bias / V & resolution / ns & delay / ns\\
			\midrule
			21 & 6.56$\pm$0.65& 28.04$\pm$0.59\\
			60 & 4.69$\pm$0.13& 23.17$\pm$0.10\\
			80 & 3.81$\pm$0.17& 22.55$\pm$0.13\\
			\bottomrule
		\end{tabular}
		\caption{\label{tab:delays}. Summary of the time resolution as well as the absolute delay between scintillator and \hb. Note that differences in the cable delays have been taken into account.}
	\end{table}

	\subsection{LGAD Timing Plane}
	
\begin{figure}[htbp]
		\centering
		\includegraphics[width=.8\columnwidth]{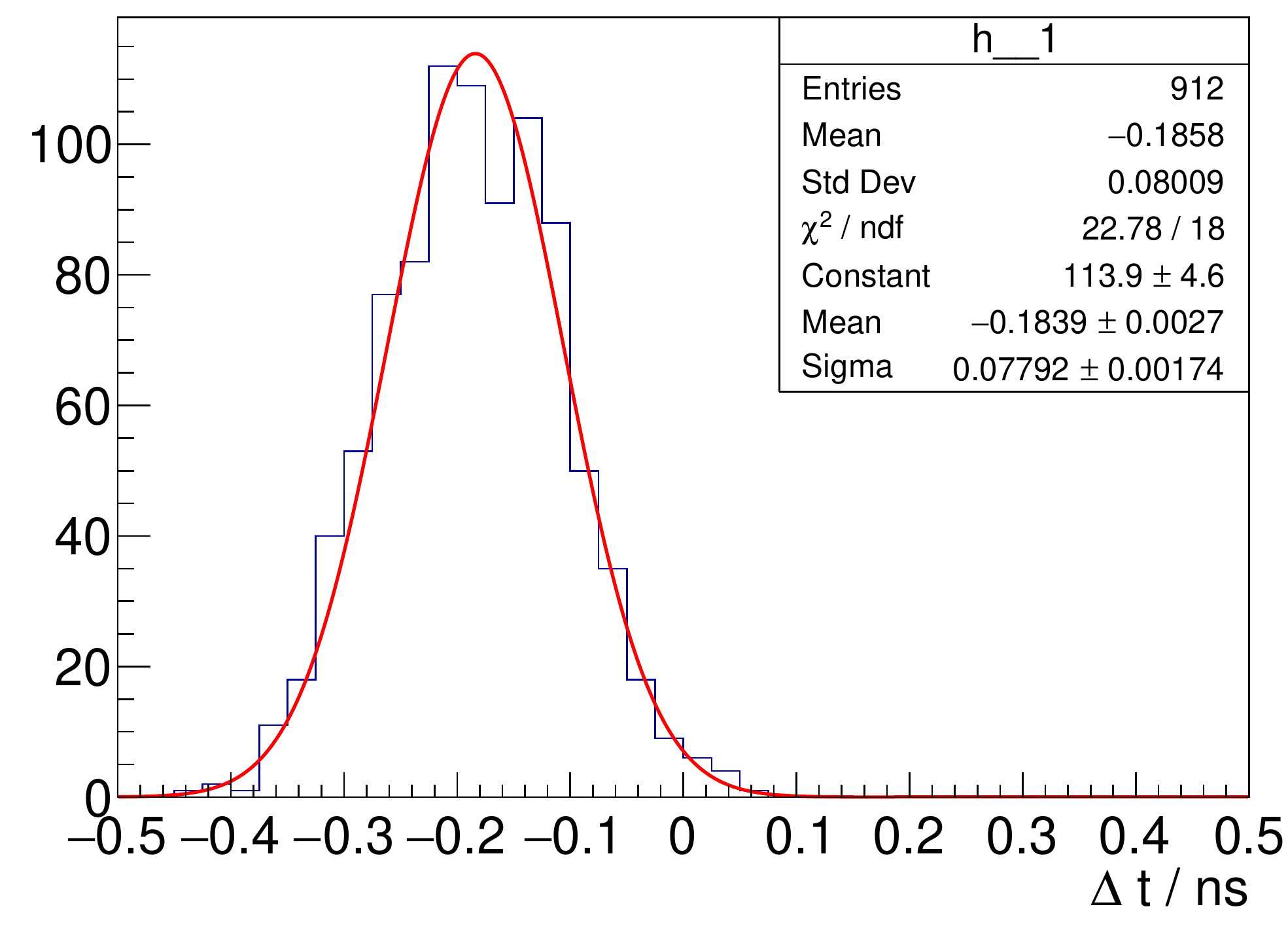}
		\caption{\label{fig:lgad_res}The measured time resolution of two LGAD layers relative to each other, measured with \SI{5.8}{GeV} electrons.}
	\end{figure}

	The proposed LGAD timing layer \cite{Sadrozinski:2014dkk,Cartiglia:2017mcj} is based on two LGADs with 5x5 pixels produced by HPK, featuring JTE separation between segments and a pitch of \SI[parse-numbers=false]{1.3x1.3}{ mm\squared}.  During a first test beam, the LGADs have been characterized to demonstrate their timing capabilities, by placing two layers behind each other and triggering on any pixel hit of the front LGAD layer. The full analog waveform has been stored on four DRS4 boards \cite{RITT2004470}, that have been synchronized to each other. The trigger ID has been sampled as an individual waveform in addition, due to the required bandwidth and buffer depth, only the six least significant bits of the ids could be recorded. The synchronization between telescope and LGADs could be verified in a recent test beam campaign. The time resolution of a single LGAD plane is extracted by the time difference  between the two layers, if both see the particle, compare figure~\ref{fig:lgad_res}. A time resolution of \SI{77.9\pm0.2}{ps} has been determined in an offline analysis for two LGADs with a bias voltage of 150V, resulting in an average individual LGAD time resolution of \SI{55}{ps}.\\
	The next generation readout, based on a 16 channel CAEN digitizer and two fast crystals as reference system  is being tested at the time of writing. 
	\section{Summary \& Outlook}
	Both telescope timing layer upgrades are well on track and show promising results:\\ 
	The \tp shows an excellent hit detection efficiency of above \SI{99.9}{\percent} at a bias voltage of \SI{-70}{V}. A time resolution of \SI{3.16\pm0.01}{ns} has been determined without any further offline corrections. The \hb logic can be used as a ROI trigger and shows small delays of down to \SI{32}{ns} with respect to a trigger scintillator.\\
	Given the very good experiences and results from the presented \tp prototype, a full scale prototype will be submitted to match the active area of the telescopes sensor planes. This is a low-risk step as the design is well advanced and takes advantage of the experience with  e.g.\ the full-size \textsc{ATLASPix3}~\cite{SCHIMASSEK2021164812} that has been designed and produced in the same \SI{180}{nm} process. \\
	The time resolution of a coincidence of two LGADs has been determined to be  \SI{77.9\pm0.2}{ps}. A more evolved and integrated DAQ system is currently being tested and installed. A full integration into the telescope DAQ system is foreseen for this year.\\
	The timing layer upgrades will enhance the performance of the EUDET-type telescopes significantly and prepares the telescope for the coming challenges of future test beam research.
	
	\section*{Acknowledgements}
					The measurements leading to these results have been performed at the Test Beam Facility at DESY Hamburg (Germany), a member of the Helmholtz Association (HGF). This project has received funding from the European Union’s Horizon 2020 Research and Innovation programme under  GA no 101004761.  We acknowledge support by the Deutsche Forschungsgemeinschaft (DFG, German Research Foundation) under Germany’s Excellence Strategy – EXC 2121 "Quantum Universe“ – 390833306. This work was supported by the UC-National Laboratory Fees Research Program grant ID \#LFR-20-653232.
	\bibliography{bib}
\end{document}